# Large-area and high-quality 2D transition metal telluride

Jiadong Zhou[1†], Fucai Liu[1†], Junhao Lin[2,3,4]*, Xiangwei Huang[5†], Juan Xia[6], Bowei Zhang[1], Qingsheng Zeng[1], Hong Wang[1], Chao Zhu[1], Lin Niu[1], Xuewen Wang[1], Wei Fu[1], Peng Yu[1], Tay-Rong Chang[7], Chuang-Han Hsu[8,9], Di Wu[8,9], Horng-Tay Jeng[7,10], Yizhong Huang[1], Hsin Lin[8,9], , Zexiang Shen[1,6,11], Changli Yang[5,12], Li Lu[5,12], Kazu Suenaga[4], Wu Zhou[2], Sokrates T. Pantelides[2,3], Guangtong Liu[5]* and Zheng Liu[1, 13,14]*.

[1]Centre for Programmable Materials, School of Materials Science and Engineering, Nanyang Technological University, Singapore 639798, Singapore

[2]Materials Science and Technology Division, Oak Ridge National Lab, Oak Ridge Tennessee 37831, USA

[3]Department of Physics and Astronomy, Vanderbilt University, Nashville, TN 37235, USA

[4]National Institute of Advanced Industrial Science and Technology (AIST), Tsukuba 305-8565, Japan

[5]Beijing National Laboratory for Condensed Matter Physics, Institute of Physics, Chinese Academy of Sciences, Beijing 100190, China

[6]Division of Physics and Applied Physics, School of Physical and Mathematical Sciences, Nanyang Technological University, Singapore 637371, Singapore

[7]Department of Physics, National Tsing Hua University, Hsinchu 30013, Taiwan

[8]Centre for Advanced 2D Materials and Graphene Research Centre, National University of Singapore, Singapore 117546

[9]Department of Physics, National University of Singapore, Singapore 117542

[10]Institute of Physics, Academia Sinica, Taipei 11529, Taiwan

[11]Centre for Disruptive Photonic Technologies, School of Physical and Mathematical Sciences, Nanyang Technological University, Singapore 637371, Singapore

[12]Collaborative Innovation Center of Quantum Matter, Beijing 100871, China

[13]Centre for Micro-/Nano-electronics (NOVITAS), School of Electrical & Electronic Engineering, Nanyang Technological University, 50 Nanyang Avenue, Singapore 639798, Singapore

[14]CINTRA CNRS/NTU/THALES, UMI 3288, Research Techno Plaza, 50 Nanyang Drive, Border X Block, Level 6, Singapore 637553, Singapore

† These authors contributed equally to this work. Correspondence and requests for materials should be addressed to J.L (email:lin.junhao@aist.go.jp), G.L (email:gtliu@iphy.ac.cn) and Z.L. (email: z.liu@ntu.edu.sg)

**Abstract**

Atomically thin transitional metal ditellurides like $WTe_2$ and $MoTe_2$ have triggered tremendous research interests because of their intrinsic nontrivial band structure. They are also predicted to be 2D topological insulators and type-II Weyl semimetals. However, most of the studies on ditelluride atomic layers so far rely on the low-yield and time-consuming mechanical exfoliation method. Direct synthesis of large-scale monolayer ditellurides has not yet been achieved. Here, using the chemical vapor deposition (CVD) method, we demonstrate controlled synthesis of high-quality and atom-thin tellurides with lateral size over 300 µm. We found that the as-grown $WTe_2$ maintains two different stacking sequences in the bilayer, where the atomic structure of the stacking boundary is revealed by scanning transmission electron microscope (STEM). The low-temperature transport measurements revealed a novel semimetal-to-insulator transition in $WTe_2$ layers and an enhanced superconductivity in few-layer $MoTe_2$. This work paves the way to the synthesis of atom-thin tellurides and also quantum spin Hall devices.

**Introduction**

Distinct from most semiconducting transition-metal sulfide and selenide compounds that are only stable in the 2H phase, tellurides exhibit much richer structural variations and electronic properties, from semiconducting in the 2H phase to semi-metallic in the 1T′ phase. Among all telluride compounds, $MoTe_2$ crystal is stable in both 2H and 1T′ phase while $WTe_2$ only form 1T′ phase in nature (Fig. 1a). Therefore, transition metal ditellurides, represented by $MoTe_2$ and $WTe_2$, have attracted tremendous interests recently due to the facile switching between different phases and their unique electronic properties. For instance, large and non-saturating magnetoresistance in $WTe_2$ bulk crystal was reported by Ali *et al*(*1*), presumably due to the perfect compensation of the electrons and holes even in an ultrahigh magnetic field. A laser-induced transition between 2H and 1T′ phases in $MoTe_2$ thin film has been demonstrated to show an Ohmic homojunction(*2*). Furthermore, high mobility up to 4000 $cm^2$/Vs and 10000 $cm^2$/Vs were reported in $MoTe_2$ and $WTe_2$, respectively (*3, 4*). $MoTe_2$ and $WTe_2$ atomic layers are also predicted to be promising candidates of type-II Weyl semimetals (*5, 6*), even in their alloys 1T′ (W, Mo)Te$_2$ (*7*). More importantly, monolayer tellurides like WTe$_2$ and MoTe$_2$ are predicted to be a 2D topological insulator(*8*). These make transition metal ditellurides an excellent platform for studying fundamental physical phenomena, such as superconductivity(*9*) and quantum spin Hall effect(*8*). They are also promising candidates for various potential applications such as spintronics and high-efficiency thermoelectric devices (*10, 11*).

However, most of the novel physical phenomena in transition-metal ditellurides have been demonstrated in mechanically exfoliated layers, e.g., the magnetoresistance in few-layer WTe$_2$(*4*) and the superconductivity in bulk 1T′-MoTe$_2$(*12*). The exfoliation method is low-yield and time-consuming and only good for scientific research. Direct synthesis of few-layer and monolayer ditelluride would, therefore, be essential to large-scale applications. Among all the growth techniques, chemical vapor deposition (CVD) has been demonstrated as a facile method in synthesizing monolayer crystals in large scale, including graphene(*13*), MoS$_2$/WS$_2$(*14, 15*), MoSe$_2$/WSe$_2$(*16, 17*), and their heterojunctions(*18-20*), by using different precursors under optimized reaction conditions. Although telluride compounds such as MoTe$_2$ films were successfully synthesized by thermal flux and tellurization of molybdenum films(*3, 21*), controlled synthesis of high-quality telluride atomic layers, even down to monolayer, remains elusive under the existing CVD or physical vapor deposition (PVD) conditions, mainly due to the lower environmental stability and activity of tellurium. For instance, the low chemical reaction activity of W and Mo with Te limits tellurization of W and Mo precursors (powder and oxides), although this method has been widely adopted for the preparation of sulfide and selenide monolayers. More specifically, the electronegativity difference between transitional metal (W or Mo) and Te is very small (~ 0.4 eV or 0.3 eV), indicating a weak bonding between the metals and Te atoms which makes the stoichiometry of ditellurides difficult to be obtained. Furthermore, even though the stoichiometry of $WTe_2$ and $MoTe_2$ is maintained, the as-synthesized product tends to decompose rapidly by emitting Te vapor at high reaction temperature (around 600 $^oC$)(*10*), instead of evaporating into gas-phase telluride.

Here, we demonstrate a CVD strategy to directly synthesize $WTe_2$ and $MoTe_2$ few-layer and monolayers at large scale. The mixed compounds (weight ratio of the compound metal oxide: metal chlorides: Te is 1:1:1) and Te were used as the source of W (Mo) and Te, respectively. Such configuration of the precursors makes the reaction between Te and the metal sources react easily. The Te in the mixed compounds can decrease the melting point of the mixed compounds, while the other Te powder in the upstream was used as to keep the Te atmosphere in the whole reaction process. The corresponding chemical reaction is shown in Supplementary Information. Raman spectroscopy was used to characterize the quality of the $WTe_2$ and $MoTe_2$ flakes to confirm their high quality. Two different stacking sequences of $WTe_2$ bilayers were observed by STEM. The as-grown $WTe_2$ atomic layers show non-saturating magnetoresistance which resembles the features of the mechanical exfoliated ones. A novel semimetal-to-insulator transition is observed in the electrical measurements of $WTe_2$ few layers, whereas the few-layer $MoTe_2$ shows enhanced superconducting behavior.

## **Results**

Side view and top view of the crystal structure of 1T′ $WTe_2$ and $MoTe_2$ is shown in Fig. 1a and 1b, respectively. A schematic diagram of the reaction system is shown in Supplementary Fig. S1. The reaction temperature is between 750 ~ 850 $^oC$ (see Methods). The thickness of the $WTe_2$ and $MoTe_2$ atomic layers can be controlled by the growth time. Single crystalline monolayer $WTe_2$ can be consistently obtained in 5 min at 820 $^oC$ with specific flow amount of carrier gas (100 sccm

Ar/15 sccm H$_2$), while increasing the reaction time results in thick WTe$_2$ flakes. Optical and atomic force microscopy (AFM) images of WTe$_2$ flakes with different thicknesses are shown in Fig. S2 and S3, respectively. Figure 1c shows a single-crystalline monolayer WTe$_2$ with a length of ~ 350 µm and width of ~ 20 µm. The sharp edges at the two ends of this rectangle suggest the edge of WTe$_2$ in 1T′ phase is terminated by planes that are intersected by 50°. Figure 1d shows the optical image of a polycrystalline WTe$_2$ monolayer film with domain size exceeding 100 µm. Figure 1e shows a bilayer WTe$_2$ with a grain boundary at the intersection of two rectangular WTe$_2$ domains (center of the image), as highlighted by the dashed line. For better visibility, the grain boundary with contrast enhancement and false color is shown in the inset of Fig. 1e. The SEM images of WTe$_2$ films are shown in Supplementary Fig. S4. Under similar conditions, ribbon-like monolayer 1T′ MoTe$_2$ up to 150 µm (Fig. 1f) and few layered 1T′ MoTe$_2$ up to 200 µm (Fig. 1g) are also obtained by controlling the Te source according to the phase diagram(*3*). Unlike WTe$_2$, different number of layers can be found in the same flake, as shown in Fig. 1g, where monolayer (1L), bilayer (2L) and trilayer (3L) MoTe$_2$ are clearly distinguished by sharp contrast. Atomic force microscopy (AFM) and SEM images of MoTe$_2$ flakes are shown in Supplementary Fig. S5 and S6, respectively.

Raman spectroscopy was employed to characterize the quality of the WTe$_2$ and MoTe$_2$ atomic layers. Figure 2a shows the Raman spectra of WTe$_2$ films with different thicknesses ranging from monolayer to bulk. For few-layer WTe$_2$ (< 10 L), only four optical vibrational modes, namely $B_1^{10}$, $A_2^3$, $A_1^7$ and $A_1^9$ were identified, compared with that reported in WTe$_2$ crystal and flakes(*22, 23*). Interestingly, the $B_1^{10}$ mode was not reported in few-layer WTe$_2$. Furthermore, the intensity of $A_1^7$ peak becomes stronger than other modes as the layer number decreases, similar to the reported result that collected the Raman spectra along *b* axis of the mechanically exfoliated WTe$_2$ atomic

layer(*24*). These results further confirm the high quality of our as-synthesized WTe$_2$ atomic layers. The optical image and Raman intensity mapping ($A_1^9$, ~ 205 cm$^{-1}$) of WTe$_2$ is shown in Fig. 2b and 2c, respectively. The region for Raman mapping is highlighted in blue dashed square and the mapping size is around 20×20 µm. Raman mapping shows homogeneous intensity across the whole region, which indicates a low defect concentration in the as-synthesized WTe$_2$ monolayers. We also find that the grain boundary in WTe$_2$ can be easily distinguished from the nearby regions in the Raman intensity mapping (Supplementary information Fig. S7). Monolayer and few-layer Raman spectra contour map of MoTe$_2$ is shown in Fig. 2d. The Raman peaks in monolayer were observed at 127, 161, 189 and 267 cm$^{-1}$, corresponding to the Raman-active A$_g$ modes of monolayer MoTe$_2$ in 1T′ phase (*25*) (Supplementary information Fig. S8). The spectrum agrees well with the previous reported result (*3, 21, 26*). The Raman spectrum of 2H and 1T′ MoTe$_2$ synthesized by controlling the Te source are shown in Supplementary information Fig. S8. Figure 2e shows a typical optical image of 1T' MoTe$_2$ few layers. Raman intensity mapping was collected at the region highlighted by the blue dashed square. Due to the high contrast, monolayer and bilayer MoTe$_2$ can be easily differentiated from Raman mapping (Fig. 2f). In addition to Raman characterization, X-ray photoelectron spectroscopy (XPS) was used to analyze the elemental distribution of WTe$_2$ and MoTe$_2$ atomic layers. The ratio of transition metals (W or Mo) and Te were found to be very close to 1:2, in good agreement with the stoichiometry of WTe$_2$ and MoTe$_2$, as shown in Supplementary Fig. S9.

Atom-resolved scanning transmission electron microscope (STEM) was applied to further investigate the atomic structure of the as-synthesized MoTe$_2$ and WTe$_2$ atomic layers. Figure 3a shows a high-resolution Z-contrast STEM image of a monolayer WTe$_2$, revealing the 1T′ phase which composes of quasi-one-dimensional tungsten-tellurium zigzag chains along the *a* axis of the unit cell (highlighted by the dashed white rectangle) and connected by Te atoms in between, as indicated in the overlaid atomic structural model. The connected Te atoms can also be viewed as a Te chain parallel to the W-Te chains. The Fast Fourier transformation (FFT) pattern shown in the inset further confirms the rectangular shape of the $WTe_2$ unit cell. Figure 3b shows a line intensity profile along the *b* axis of the crystal, indicating two distinct positions for Te atoms bonding to the W atom in the distorted 1T phase, with a measured distance of 2.49 Å and 1.61 Å, respectively. The simulated STEM image using the overlaid atomic structural model achieves a good agreement with the experimental image, as shown in the inset of Fig. 3a. TEM characterization on thick flakes $WTe_2$ is shown in Supplementary Fig. S10, further confirming the high quality of the sample.

The as-synthesized $MoTe_2$ atomic layer also maintains the 1T′ phase similar to the $WTe_2$ layer presented above, as shown in Supplementary Fig. S11, where the electron-energy-loss spectra (EELS) are provided for direct comparison between the two materials. In light of the co-existence of different stacking orders in 2D materials, we also found that the CVD-grown 1T′ $WTe_2$ maintains two different stacking sequences in the bilayer that are similar to other 2D materials. Fig. 3c and 3d show the atomic structures of two $WTe_2$ bilayer regions, which are distinguishable from each other. Specifically, Figure. 3c shows a bilayer stacking where the second layer is mirror symmetric to the first layer along the *b* axis of the $WTe_2$ unit cell, as shown by the opposite orientations of the two layers in the top view of the structural model (the dashed diamonds). The Te chains of the second layer aligns with the W-Te zigzag chains of the first layer vertically, and vice versus, which can be visualized in the side view of the structural model. Such kind of stacking is similar to the 2H stacking in TMDs and AA' stacking in h-BN. Therefore, we called it 2H stacking following the convention. Figure 3d shows another stacking pattern of $WTe_2$, where the second layer shifts half of a unit cell along the *b* axis away from the 2H stacking shown in Fig. 3c, as indicated by the structural model. We called such stacking 2H′ stacking. Both stackings form periodic stripe patterns, which consist of overlapped Te and W-Te chains from the two layers, in good agreement with the simulated STEM images (Supplementary information Fig. S12). Density functional theory (DFT) calculations on the 2H and 2H′ stacking show very similar band structure, both of which maintain semi-metallic properties, as shown in Supplementary Fig. S13. Interlay shifting along *b* axis(*27*), however, is rarely observed in WTe$_2$, presumably due to the large interlayer interaction in materials composed by heavy elements. Figure 3e shows an atomically sharp stacking boundary between the 2H and 2H′ stacking domains, as highlighted by the green dashed rectangle. A simulated STEM image based on the DFT-relaxed model between the two stackings is shown in supplementary Fig. S13, which resembles most feature of the stacking boundary observed experimentally. Due to the symmetry of the zigzag chains, the second layers in the two stacking form are also mirror symmetric to each other which forms a mirror twin boundary, similar to the graphene domain wall(*28*). DFT calculations further indicate the Te$_2$ columns are misaligned in the relaxed structure of the mirror twin boundary (inset in Fig. 3f), different from those observed in other TMDs(*29-31*). A projected density of states (DOS) illustrates (Fig. 3f) that the W atoms in the mirror twin boundary show more states at the Fermi level, suggesting they are more metallic than the bulk counterpart, which may have important contributions in the electrical behavior to the monolayer and few-layer WTe$_2$ and also implications on the quantum spin Hall effect in WTe$_2$.

Layered CVD-grown ditelluride films provide an excellent platform to study the thickness-dependent electric transport. The electrical properties of the as-synthesized WTe$_2$ and MoTe$_2$ flakes with different thickness were investigated by means of field- and temperature-dependent transport measurements, as shown in Fig. 4. Figure 4a shows the resistivity of $WTe_2$ as a function of temperature with thickness of 4 nm (5 layer), and 2 nm (2 layer). The corresponding optical and AFM images of the $WTe_2$ devices are shown in Supplementary Fig. S14. Above 50 K, the resistance of 4 nm $WTe_2$ decreasing monotonically with decreasing temperature shows a metallic behavior, which is similar to the transport properties of the bulk $WTe_2$, However, an upturn in resistance curve was observed with $T$ further reduced. Such phenomenon may be due to the two-dimensional electron-electron interactions at the reduced dimension (*32*). Interestingly, the bilayer $WTe_2$ (with thickness of 2 nm) displays solely insulating behavior under zero magnetic field, confirming that the semimetal-to-insulator transition originates from the effect of the reduced thickness. Such phenomenon has been observed in mechanically exfoliated $WTe_2$ few layers, where the insulating state may be attributed to the Anderson localization in the two dimensional limit(*4*). The temperature-dependent magnetoresistance (MR) calculated by MR=[$\rho(H)$-$\rho(0)$]/$\rho(0)$ of bilayer $WTe_2$ as a function of the magnetic field is shown in Fig. 4b. These results show that large and non-saturating magnetoresistance is preserved in our CVD-grown $WTe_2$ even down to a bilayer sample, which further demonstrates their high quality. The MR reaches a maximum value of 28% at 2 K. For the thick $WTe_2$ flakes (12 nm), the MR is about 2000% at 25K in a field of 10T, which is shown in Supplementary Fig. S15. These values are close to the order of magnitude of the recent reports(*4*).

Enhanced superconductivity is also observed in our as-synthesized few-layer $MoTe_2$. The optical image of the $MoTe_2$ device is shown in Supplementary Fig. S16. Figure 4c shows the longitudinal resistance $R_{xx}$ as a function of temperature $T$ of few-layer $MoTe_2$ device in different perpendicular magnetic fields. $R_{xx}$ decreases steadily from 300K to 40K, indicating that the sample shows a metallic behavior which is consistent with previous studies on exfoliated samples (*3*). With $T$ further reduced, the sample gradually becomes superconducting below $T$=2.5K (the onset of transition) and reaches zero resistance at $T_c$=0.5K. Surprisingly, the superconductivity in thinner samples is strongly enhanced compared with $T_c$=0.1K reported in its bulk counterpart (*12*). A similar phenomenon was observed in 2H TaS$_2$(*33*), and the reason is ascribed to an enhancement of the effective electron–phonon coupling constant in thinner samples. The inset of Fig. 4c displays the longitudinal resistance $R_{xx}$ as a function of temperature in different perpendicular magnetic fields. We define the superconducting transition temperature $T_c$ under different magnetic fields as the temperature at which the resistance drops to 10% of the normal state resistance $R_N$. $T_c$ shifts systematically to lower temperatures with increasing magnetic fields $B$. Finally, the superconductivity was completely suppressed when $B \geq 1$T. We summarize the upper critical field $H_{c2}$-$T_c$ phase diagram in Fig. 4d and find a linear relationship between $H_{c2}$ and $T_c$ near $T_c$. This is a characteristic of 2D superconductors and can be explained by the standard linearized Ginzburg-Landau (GL) theory,

$$H_{c2}(T) = \frac{\phi_0}{2\pi \xi_{GL}(0)^2}\left(1 - \frac{T}{T_c}\right)$$

where $\xi_{GL}(0)$ is the zero-temperature GL in-plane coherence length and $\phi_0$ is the magnetic flux quantum. By fitting the experimental data with the above formula, a coherence length of 38 nm was obtained, which is much larger than that in Mo$_2$C(*34*).

**Discussion**

In summary, large-scale and atom-thin ditellurides including WTe$_2$ and MoTe$_2$ were synthesized. Complementary characterizations demonstrated the high quality of as-grown samples. High-resolution STEM imaging resolved that atomic structure of WTe$_2$ and MoTe$_2$ and also observed the domain wall in bi-layer WTe$_2$ where the stacking boundary are revealed between two distinct stacking sequences. Electric transport measurement also revealed the semimetal-to-insulator transition in WTe$_2$ and enhanced superconductivity in MoTe$_2$. Our work will shed light on the synthesis of atom-thin telluride materials and boost the realization of quantum spin Hall devices.

**Materials and Methods**

**Synthesis of MoTe$_2$ and WTe$_2$**. The WTe$_2$ (MoTe$_2$) crystals were synthesized by CVD method using WO$_3$ (MoO$_3$) and WCl$_6$ (MoCl$_5$) (Sigma) as the W sources. The Te powder was used as the Te sources. The crystals are synthesized in quartz tube (1 inch diameter) with temperature from 700 to 850 ºC. The system of the reaction is shown in Supplementary Fig. S1. Specifically, for the WTe$_2$, the mixed gas of H$_2$/Ar with 15 sccm and 150 sccm were used as the carrier gas, the silicon boat contained 30 mg mixed powders with WO$_3$: WCl$_6$: Te=1:1:1(weight ratio) was put in the center of the tube. The SiO$_2$/Si substrate is placed downstream. Another silicon boat containing 0.5 g Te powder is put on the upstream. The temperature ramps up to the 820 ºC in 17 min, and keeping at the reaction temperature for about 5 min to 15 min. Then the furnace cools down to room temperature gradually. For the MoTe$_2$, the mixed gas of H$_2$/Ar with 15 sccm and 200 sccm were used as the carrier gas, the silicon boat contained 30mg mixed powder with MoO$_3$: MoCl$_5$: Te=1:1:1 was put in the center of the tube. The SiO$_2$/Si substrate is placed downstream. Another silicon boat contained 0.5 g Te powders is put on the upstream. The temperature ramps up to the 780 ºC in 16 min, and keeping at the reaction temperature for about 5 min to 15 min. Then the furnace cools down to room temperature gradually. Detailed description of the growth is given in supporting Supplementary Information.

**Raman Characterization.** Raman measurements with an excitation laser of 532 nm was performed using a WITEC alpha 200R Confocal Raman system. Before Raman characterization, the system was calibrated with the Raman peak of Si at 520 cm$^{-1}$. The laser powers are less than 1mW to avoid overheating of the samples.

**TEM and STEM Characterization.** The STEM samples were prepared with a poly (methyl methacrylate) (PMMA) assisted method. A layer of PMMA of about 1 µm thick was spin-coated on the wafer with $WTe_2$ ($MoTe_2$) samples deposited, and then baked at 180 °C for 3min. Afterwards, the wafer was immersed in NaOH solution (1M) to etch the $SiO_2$ layer over night. After lift-off, the PMMA/$WTe_2$ ($MoTe_2$) film was transferred into DI water for several cycles to wash away the residual contaminants, and then it was fished by a TEM grid (Quantifoil Mo grid). The transferred specimen was dried naturally in ambient environment, and then dropped into acetone overnight to wash away the PMMA coating layers. The STEM imaging shown in the main text was performed on an aberration-corrected Nion UltraSTEM-100 operating at 100 kV. The convergence semi-angle for the incident probe was 31 mrad. Z-contrast images were collected for a half-angle range of ~86-200 mrad. STEM imaging and EELS analysis on $MoTe_2$ shown in the SI were performed on a JEOL 2100F with a cold field-emission gun and an aberration corrector (the DELTA-corrector) operating at 60 kV. A Gatan GIF Quantum was used for recording the EELS spectra. The inner and outer collection angles for the STEM image (β1 and β2) were 62 and 129–140 mrad, respectively, with a convergence semi-angle of 35 mrad. The beam current was about 15 pA for the ADF imaging and EELS chemical analyses. All imaging was performed at room temperature.

**Devices fabrication and transport measurement.** The Hall bar are patterned on few layer $MoTe_2$ and $WTe_2$ using e-beam lithography. The Ti/Au (5/50 nm) electrodes are deposited using the thermal evaporator, followed by the lift off process. For the measurements of WTe$_2$, the transport measurement is performed in the Quantum Design PPMS system with temperature ranging from 300 K to 2 K, and magnetic field up to 14 T. For the superconductivity measurements of MoTe$_2$, the transport experiment is carried out in a top-loading Helium-3 cryostat in a superconducting magnet. An ac probe current $I_{ac}$=10 nA at 30.9 Hz is applied from the source to the drain. Then a lock-in amplifier monitors the longitudinal $R_{xx}$ through two additional electrical contacts.

**Supplementary Materials**

Fig. S1. Synthesis of WTe$_2$ and MoTe$_2$ films
Fig. S2. Optical images of WTe$_2$ flakes with different thickness
Fig. S3. AFM measurements of WTe$_2$ flakes
Fig. S4. SEM images of WTe$_2$ films
Fig. S5. AFM measurement of monolayer MoTe$_2$
Fig. S6. Morphologies of monolayer, bilayer and trilayer MoTe$_2$
Fig. S7. The Raman mapping of WTe$_2$
Fig. S8. Raman spectra of 2H and 1T' MoTe$_2$
Fig. S9. XPS test of the monolayer WTe$_2$ and MoTe$_2$
Fig. S10. TEM characterize of WTe$_2$ flakes
Fig. S11. Atomic structure of MoTe$_2$ and WTe$_2$ few layers
Fig. S12. Simulated STEM image of the stacking boundary in WTe$_2$ bilayer
Fig. S13. Crystal and band structures of monolayer and bilayer WTe$_2$
Fig. S14. AFM images of the WTe$_2$ flakes used in the devices
Fig. S15. Magnetoresistance of thick WTe$_2$ flakes
Fig. S16. Optical image of MoTe$_2$ device

**Acknowledgements**

This work is supported by the Singapore National Research Foundation under NRF RF Award No. NRF-RF2013-08, the start-up funding from Nanyang Technological University (M4081137.070). JL and KS acknowledge support from the JST Research Acceleration Programme. This research was also supported in part by U.S. DOE grant DE-FG02-09ER46554 (JL, STP), by the U.S. Department of Energy, Office of Science, Basic Energy Science, Materials Sciences and Engineering Division (WZ), and through a user project at ORNL's Center for Nanophase Materials Sciences (CNMS), which is a DOE Office of Science User Facility. This research used resources of the National Energy Research Scientific Computing Center, which is supported by the Office of Science of the US Department of Energy under Contract No.DE-AC02-05CH11231. T.R.C. and H.T.J. are supported by the Ministry of Science and Technology, National Tsing Hua University, and Academia Sinica, Taiwan. We also thank NCHC, CINC-NTU and NCTS, Taiwan for technical support. H.L. acknowledges the Singapore National Research Foundation for the support under NRF Award No. NRF-NRFF2013-03. The work at IOP was supported by the National Basic Research Program of China from the MOST under the grant No.2014CB920904 and 2013CB921702, by the NSFC under the grant No. 11174340 and 91421303.

Additional information competing financial interests: The authors declare no competing financial interests.

**Author contributions**

J.Z., Z.L. designed the experiments. J.Z worked on the growth of $WTe_2$ and $MoTe_2$. J.X. and J.Z. carried out Raman characterizations and analyzed the data. J.L. performed all the STEM characterization of the samples and analysis. T.R.C., C.-H.C, D.W., H.T.J, and H.L. performed electronic structure calculations. B,Z. performed the TEM of thick $WTe_2$ flakes. J.Z. performed the AFM test. F.L fabricated and measured the $WTe_2$ devices. Q,Z. and C,Z. fabricated the $MoTe_2$ devices. X.H. measured the superconductivity in $MoTe_2$. J.Z., J.L. and Z.L. wrote the paper. All authors discussed and commented on the manuscript.

# Figure and Figure Caption

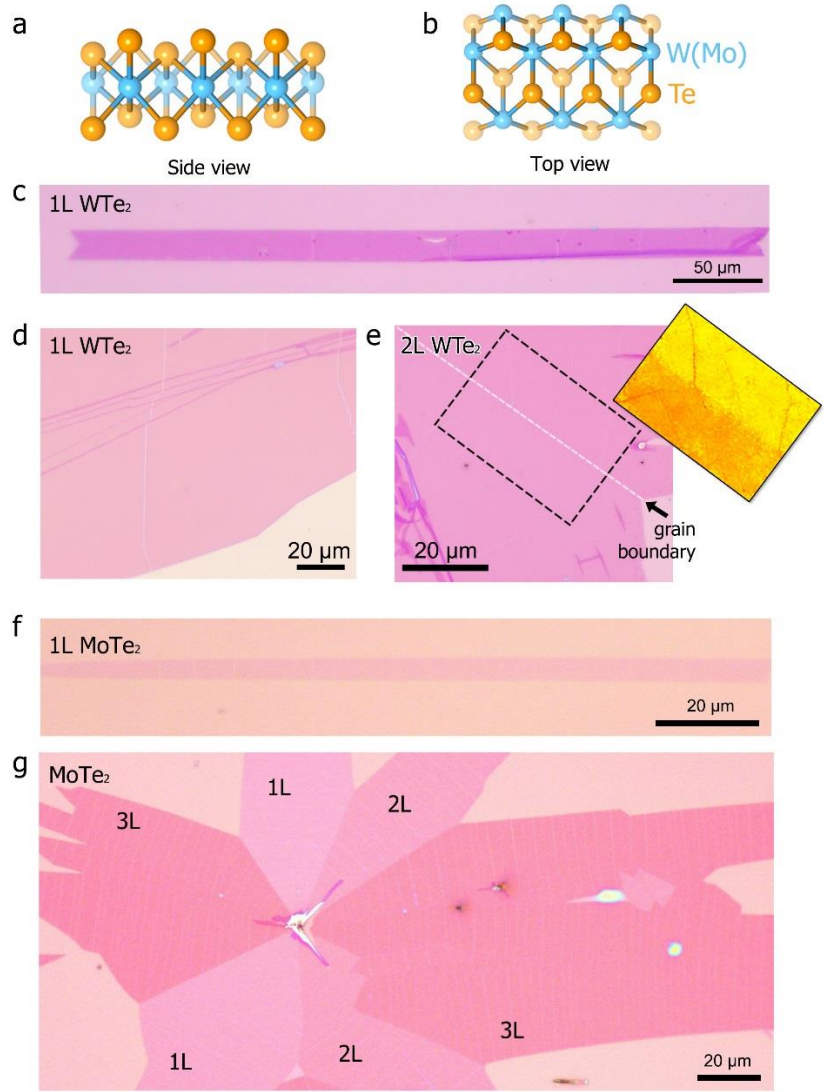

**Fig.1. Optical geometries of WTe$_2$ and MoTe$_2$ monolayers. a**, **b**, Side and top views of the crystal structure of 1T' W(Mo)Te$_2$, respectively. **c**, Optical image of a large single crystalline WTe$_2$ monolayer with a length of ~ 350 µm and width of ~ 20 µm. **d,** Optical image of a large WTe$_2$ monolayer film. **e**, Optical image of a large bilayer WTe$_2$ with grain boundary. The grain boundary is indicated by the arrow and the dashed line. Inset: false color image of the region indicated by the dashed rectangle, highlighting the location of the grain boundary. **f**, Optical image of a single crystalline MoTe$_2$ monolayer. With the length and width of ~ 150 and 8 µm, respectively. **g**, Optical image of a MoTe$_2$ flake containing 1L, 2L and 3L MoTe$_2$. The number of layer can be easily identified by their contrast.

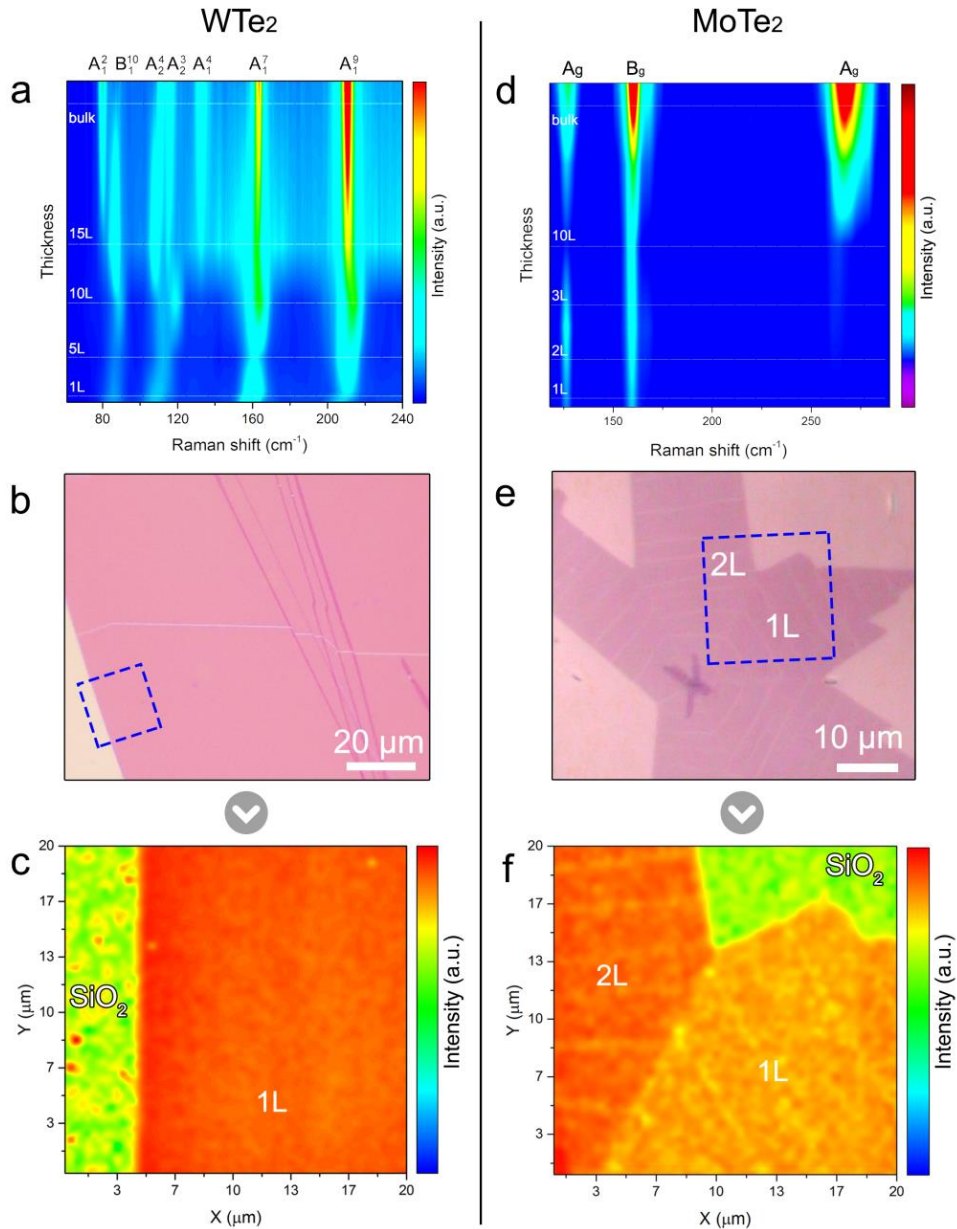

**Fig.2. Raman spectra and mapping of WTe$_2$ and MoTe$_2$ monolayers. a**, Thickness-dependent Raman spectra of WTe$_2$. Seven characteristic vibration modes are identified in bulk WTe$_2$ while only 4 of them can be observed in few-layer WTe$_2$. **b**, Optical image of a WTe$_2$ monolayer. **c**, Raman intensity mapping (from $A_1^9$) from the region (highlighted by blue dashed square) in **b**. The mapping size is 20 × 20 µm. The left side is SiO$_2$ while the right side shows a homogenous WTe$_2$ film. **d**, Thickness-dependent Raman spectra of MoTe$_2$. Four characteristic vibration modes are identified in bulk MoTe$_2$. Raman intensity dramatically decreases for all Raman peaks in MoTe$_2$ thin flakes. **e**, Optical image of MoTe$_2$ containing monolayer and bilayer regions. **f**, Raman intensity mapping (A$_g$@161 cm$^{-1}$) from the region highlighted by blue dashed square in **e**. The mapping size is 20 ×20 µm.

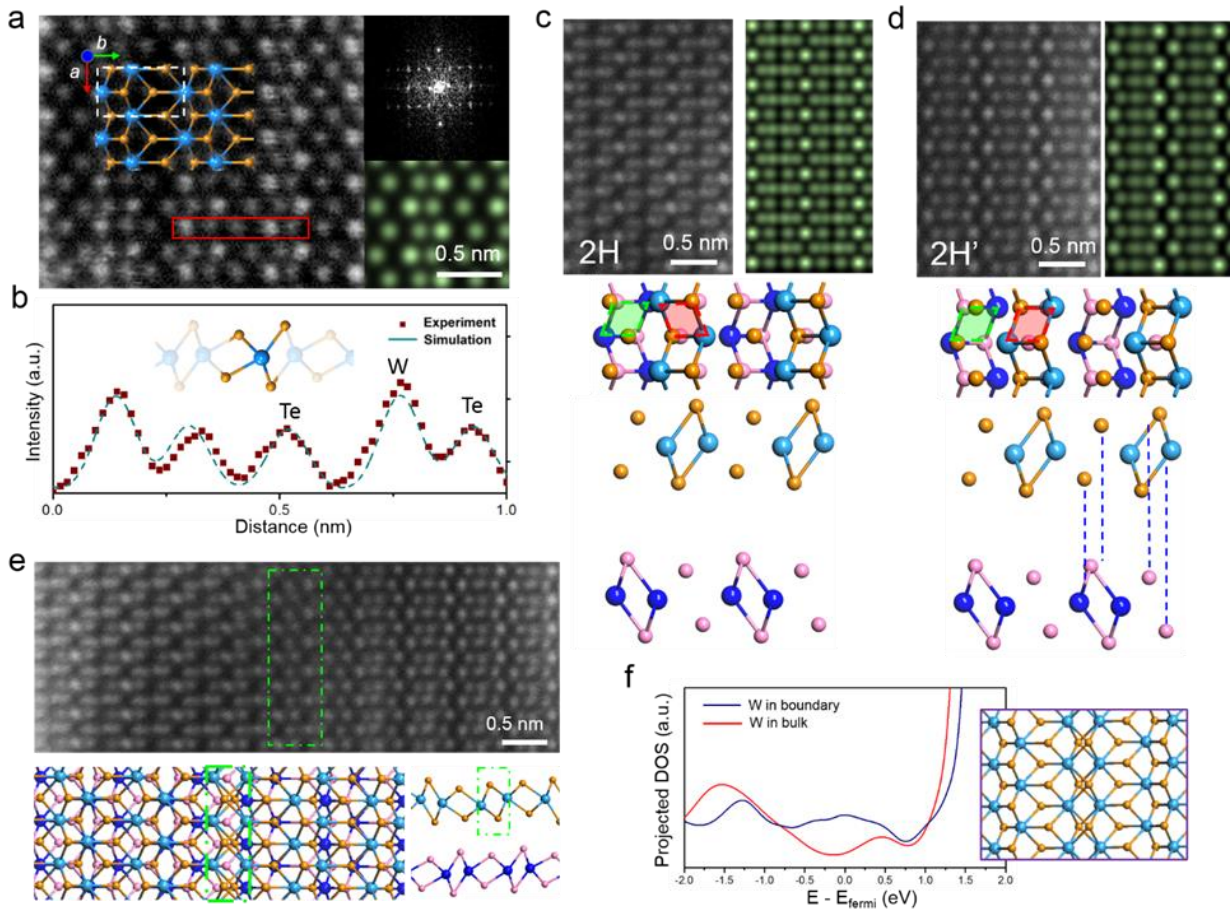

**Fig.3. Atomic resolution STEM characterization of monolayer and bilayer WTe$_2$. a,** STEM Z-contrast image of a monolayer WTe$_2$. The coordinate and structural model are overlaid on the image. Insets: FFT pattern and simulated STEM image of the monolayer WTe$_2$. **b,** Line intensity profile file of the region highlighted by red rectangle in **a**. **c, d,** STEM Z-contrast image of a bilayer WTe$_2$ with 2H stacking **c** and 2H' stacking **d**. The green and red dash diamonds indicate the orientation of the zigzag W-Te chains in the first and second layer, respectively. 2H and 2H′ stacking is differed by half of a unit cell shifting along the *b* aixs in the second layer. Images in green are simulated images. **e,** STEM Z-contrast image of an atomically sharp stacking boundary between the 2H (left) and 2H′ (right) stacking. The structural model is optimized by DFT calculations. **f,** Projected DOS of the W atoms in the mirror twin boundary and in the bulk, showing more states near the Fermi level for W atoms in the boundary region. Inset: the structural model of the mirror twin boundary.

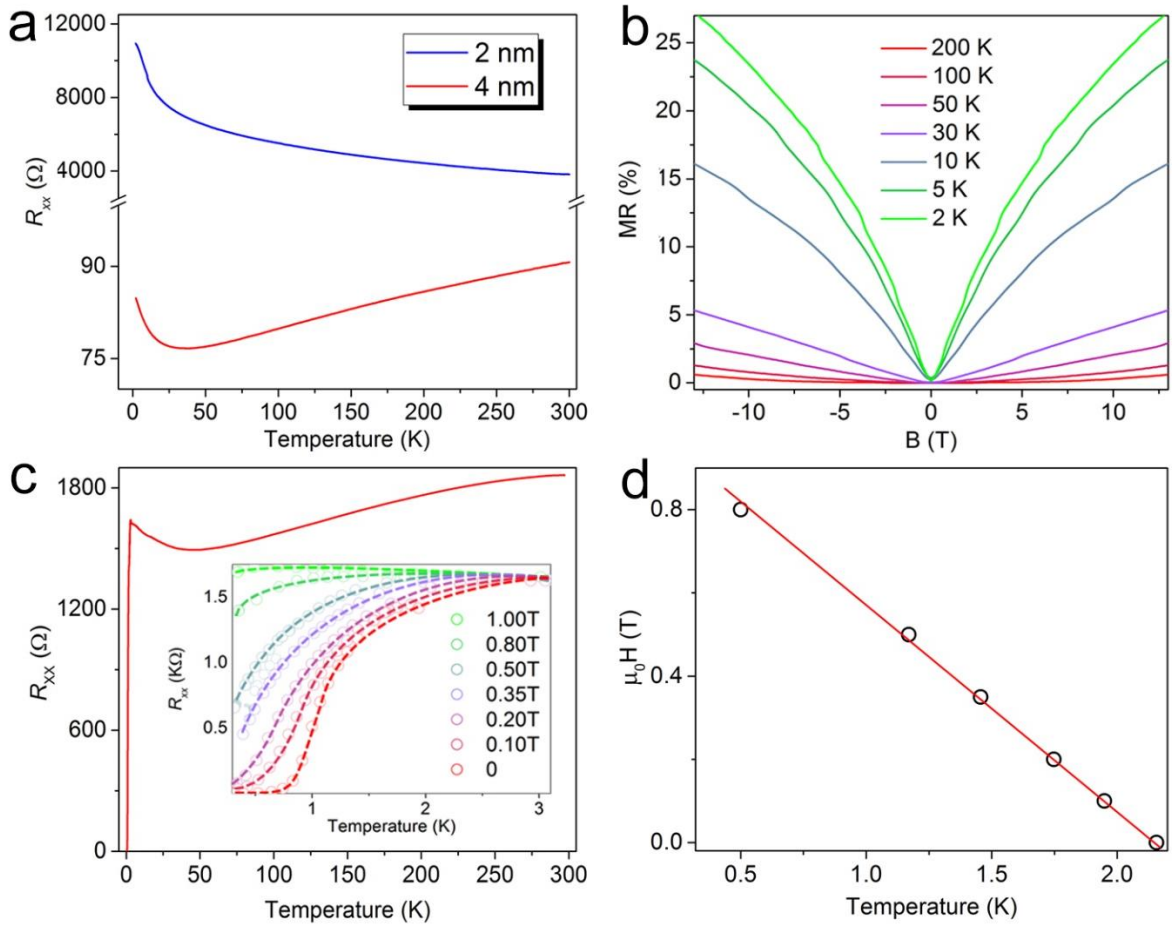

**Fig.4. Transport in different thicknesses of WTe$_2$ and superconductivity in few layered MoTe$_2$. a**, Temperature-dependent resistance of WTe$_2$ flakes with a thickness of 4 nm and 2 nm under zero magnetic field, respectively. The corresponding field dependent magnetoresistance of 2 nm WTe$_2$ flake at different temperatures is shown in **b**. WTe$_2$ shows a semimetal-to-insulator transition as the thickness decreased from 4 nm to 2 nm. **c**, Superconducting resistive transition of a few layered MoTe$_2$ at zero magnetic field. Inset: Superconductivity of the sample in different perpendicular magnetic fields. **d**, Temperature dependence of the upper critical field $H_{c2}$. The solid red line is the linear fit to $H_{c2}$.